\documentclass[sigconf]{acmart}
\usepackage{adjustbox}
\usepackage{algorithmic}
\usepackage{graphicx}
\usepackage{textcomp}
\usepackage{xcolor}
\usepackage{bbding}
\usepackage{tabulary}
\usepackage{booktabs}%
\usepackage{siunitx}
\usepackage[acronym]{glossaries}
\usepackage[font={footnotesize}]{caption}
\usepackage{subcaption}
\usepackage{url}
\usepackage{comment}
\usepackage[table]{xcolor}
\usepackage{balance}
\newacronym{swe}{SWE}{Snow Water Equivalent}
\newacronym{fmcw}{FMCW}{Frequency-Modulated Continuous-Wave}
\newacronym{gfrp}{GFRP}{Glass-Fibre Reinforced Plastic}
\newacronym{saw}{SAW}{Surface Acoustic Wave}
\newacronym{fscw}{FSCW}{Frequency Stepped Continuous Wave}
\newacronym{rtd}{RTD}{Resistance Temperature Detectors}
\newacronym{mmwave}{mmWave}{Millimeter-Wave}
\newacronym{if}{IF}{Intermediate Frequency}
\newacronym{fft}{FFT}{Fast-Fourier Transform}
\newacronym{dc}{DC}{Direct Current}
\newacronym{iq}{IQ}{In-phase and Quadrature}
\newacronym{ml}{ML}{Machine Learning}
\newacronym{ap}{AP}{Amplitude Profile}
\newacronym{tcap}{TCAP}{Temperature Compensated Amplitude Profile}
\newacronym{tsr}{TSR}{Temperature Sensor Reference}

\def\BibTeX{{\rm B\kern-.05em{\sc i\kern-.025em b}\kern-.08em
    T\kern-.1667em\lower.7ex\hbox{E}\kern-.125emX}}

\AtBeginDocument{%
  \providecommand\BibTeX{{%
    Bib\TeX}}}

\copyrightyear{2025}
\acmYear{2025}
\setcopyright{cc}
\setcctype{by}
\acmConference[UbiComp Companion '25]{Companion of the the 2025 ACM International Joint Conference on Pervasive and Ubiquitous Computing}{October 12--16, 2025}{Espoo, Finland}
\acmBooktitle{Companion of the the 2025 ACM International Joint Conference on Pervasive and Ubiquitous Computing (UbiComp Companion '25), October 12--16, 2025, Espoo, Finland}\acmDOI{10.1145/3714394.3756221}
\acmISBN{979-8-4007-1477-1/2025/10}

\begin{document}

\title{Adaptive Internal Calibration for Temperature-Robust mmWave FMCW Radars}

%%
%% The "author" command and its associated commands are used to define
%% the authors and their affiliations.
%% Of note is the shared affiliation of the first two authors, and the
%% "authornote" and "authornotemark" commands
%% used to denote shared contribution to the research.
\author{Dariush Salami}
\affiliation{%
  \institution{Aalto University}
  \city{Espoo}
  \country{Finland}}
\email{dariush.salami@aalto.fi}

\author{Nima Bahmani}
\affiliation{%
  \institution{Aalto University}
  \city{Espoo}
  \country{Finland}}
\email{nima.bahmani@aalto.fi}

\author{Hüseyin Yiğitler}
\affiliation{%
  \institution{Aalto University}
  \city{Espoo}
  \country{Finland}}
\email{yusein.ali@aalto.fi}

\author{Stephan Sigg}
\affiliation{%
  \institution{Aalto University}
  \city{Espoo}
  \country{Finland}}
\email{stephan.sigg@aalto.fi}

%%
%% By default, the full list of authors will be used in the page
%% headers. Often, this list is too long, and will overlap
%% other information printed in the page headers. This command allows
%% the author to define a more concise list
%% of authors' names for this purpose.
% \renewcommand{\shortauthors}{Bahmani et al.}
%\renewcommand{\shortauthors}{Blinded for review}

%%
%% The abstract is a short summary of the work to be presented in the
%% article.

\begin{abstract}
We present a novel internal calibration framework for \gls{mmwave} \gls{fmcw} radars to ensure robust performance under internal temperature variations, tailored for deployment in dense wireless networks. Our approach mitigates the impact of temperature-induced drifts in radar hardware, enhancing reliability. We propose a temperature compensation model that leverages internal sensor data and signal processing techniques to maintain measurement accuracy. Experimental results demonstrate improved robustness across a range of internal temperature conditions, with minimal computational overhead, ensuring scalability in dense network environments. The framework also incorporates ethical design principles, avoiding reliance on sensitive external data. The proposed scheme reduces the Pearson correlation between the amplitude of the \gls{if} signal and internal temperature drift up to \qty{84}{\percent}, significantly mitigating the temperature drift.
\end{abstract}

%%
%% The code below is generated by the tool at http://dl.acm.org/ccs.cfm.
%% Please copy and paste the code instead of the example below.
%%


%%
%% Keywords. The author(s) should pick words that accurately describe
%% the work being presented. Separate the keywords with commas.
%% A "teaser" image appears between the author and affiliation
%% information and the body of the document, and typically spans the
%% page.

%\begin{teaserfigure}
%  \includegraphics[width=\textwidth]{sampleteaser}
%  \caption{Seattle Mariners at Spring Training, 2010.}
%  \Description{Enjoying the baseball game from the third-base
%  seats. Ichiro Suzuki preparing to bat.}
%  \label{fig:teaser}
%\end{teaserfigure}

%%
%% This command processes the author and affiliation and title
%% information and builds the first part of the formatted document.
\maketitle

\section{Introduction}
\gls{mmwave} \gls{fmcw} radars are pivotal in dense wireless networks, enabling applications such as environmental monitoring, structural health assessment, human activity recognition~\cite{palipana2021pantomime,salami2022tesla, salami2024spectrum, salami2023joint, salami2024angle}, and emerging biomedical applications such as non-invasive glucose monitoring~\cite{nikandish2024contactless} and arterial pulse detection~\cite{bahmani2025non}. These radars offer high resolution, penetration through non-metallic materials, and robustness in varied conditions, making them ideal for non-contact sensing. However, internal temperature variations within radar hardware, caused by prolonged operation, ambient environmental changes, or power dissipation, introduce significant measurement errors. These errors manifest as frequency drifts, phase shifts, and amplitude variations, compromising the accuracy and reliability of radar measurements~\cite{toledo2020absolute, machado2018automotive}. In dense wireless networks, where multiple devices operate concurrently, such errors can degrade system performance, disrupt data integrity, and undermine trust in critical applications.

The need to mitigate temperature-induced drifts is particularly acute in applications requiring high precision, such as non-invasive glucose monitoring~\cite{nikandish2024contactless,omer2020blood}. Glucose monitoring using mmWave radars relies on detecting subtle changes in the dielectric properties of tissue, which are influenced by glucose concentration. Temperature variations in radar hardware can mask these subtle signals, leading to inaccurate glucose readings that could have severe implications for patient health, such as misinformed insulin dosing. For instance, a frequency drift of even a few megahertz due to a \qty{10}{\celsius} temperature change can alter the phase response enough to distort the measured dielectric properties, resulting in errors exceeding acceptable clinical thresholds (e.g., \qty{10}{\milli\gram\per\deci\litre} for glucose levels). 

Wang et al.~\cite{wang2024influences} demonstrate that internal temperature variations cause significant nonlinearity in \gls{fmcw} signals, degrading radar performance. This exacerbates frequency detuning and underscores the need for our proposed internal calibration framework to mitigate temperature-induced drifts, ensuring reliable measurements in applications like structural health monitoring, where false positives in defect detection can compromise infrastructure safety, and non-invasive biomedical sensing, where precision is critical for patient outcomes.

Traditional calibration methods often rely on external reference sources or environmental sensors, which introduce complexity, increase power consumption, and raise privacy concerns in dense wireless networks. For example, in glucose monitoring, external sensors might inadvertently collect sensitive physiological data, violating ethical design principles.
In this paper, we propose an internal calibration framework that compensates for temperature-induced drifts using only onboard sensor data, aligning with the privacy-scalable and ethical design principles. By ensuring robust radar performance without external dependencies, our approach supports applications like glucose monitoring, where precision, reliability, and user privacy are paramount.

Our contributions are
\begin{itemize}
    \item an internal calibration framework for \gls{mmwave} \gls{fmcw} radars to compensate for temperature-induced drifts.
    \item a temperature compensation model using internal sensor data, ensuring privacy-scalable and ethical operation.
    \item We validate the framework’s robustness and scalability through experiments, demonstrating its suitability for dense wireless networks and critical applications like glucose monitoring.
\end{itemize}

\section{Related work}
The impact of temperature variations on \gls{mmwave} \gls{fmcw} radar performance has been studied in various contexts, particularly for structural health monitoring and environmental sensing. Simon et al. investigated temperature effects on \qty{60}{\giga\hertz} radar signals reflected by \gls{gfrp} plates, noting significant phase and amplitude variations due to temperature-induced changes in material properties and air propagation \cite{simon2021experimental}. While their work focused on external environmental factors, it highlights the need for calibration to address temperature effects.

Schuster et al. explored \gls{saw}-based temperature measurement with \gls{fmcw} radars, achieving high accuracy by compensating for thermal expansion effects in sensors \cite{schuster2006performance}. However, their approach required external reference signals, which may not be feasible in privacy-sensitive dense networks. Baer et al. used dielectric mixing equations to detect temperature-induced changes in gas permittivity, but their method relied on external environmental data, posing potential privacy risks \cite{baer2014contactless}.

Calibration techniques for radars typically involve external references, such as known targets or environmental sensors, which can introduce complexity and privacy concerns in dense wireless networks. Our work addresses these gaps by developing an internal calibration framework that uses only onboard sensor data, ensuring both robustness against temperature variations and compliance with privacy-scalable and ethical design principles.

\section{Methodology}
\begin{figure*}
    \centering
    \includegraphics[width=1\linewidth]{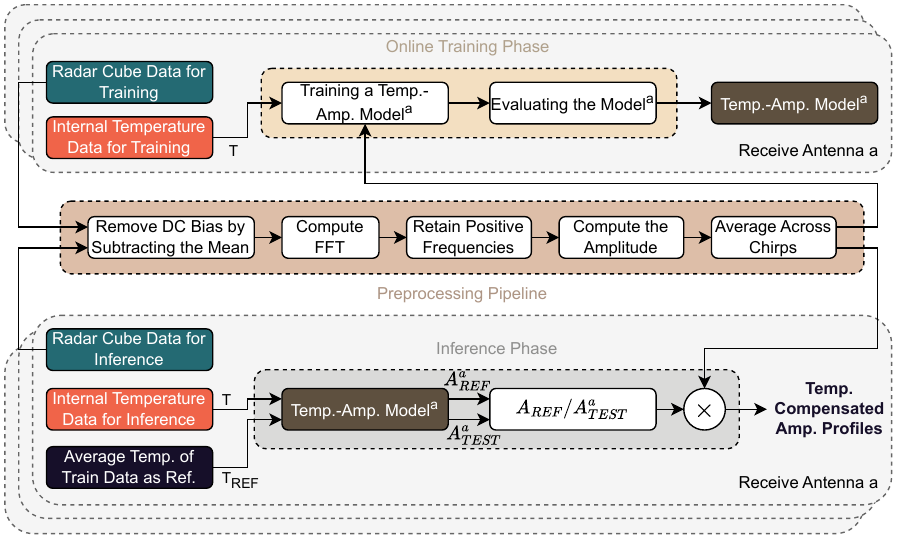}
    \caption{Block diagram of the radar system architecture, detailing the signal processing pipeline for temperature-based amplitude calibration and data analysis. The diagram employs color-coding to distinguish functional components: green represents radar signal acquisition and preprocessing, pink dashed line in the middle denotes \gls{fft}-based frequency-domain analysis, and red indicates internal temperature measurement. Received signals from multiple antennas, denoted by index $k$, are processed independently to capture amplitude displacements, enabling per-antenna calibration to mitigate temperature-induced variations in signal amplitude for enhanced measurement accuracy.}
    \label{fig:system_model}
\end{figure*}

This section outlines the radar signal preprocessing pipeline and the temperature-based amplitude calibration framework designed to mitigate the gain drift caused by internal temperature variations. The system diagram in Fig.~\ref{fig:system_model} illustrates the full flow, which is divided into three phases: Preprocessing, Online Training, and Inference.

\subsection{Preprocessing Pipeline}
Each radar frame consists of complex baseband \gls{iq} samples collected across multiple chirps and samples per chirp for each antenna. Let the radar data cube be denoted as

\begin{equation}
    \mathcal{R} \in \mathbb{C}^{F \times A \times C \times N},
\end{equation}

\noindent where $F$ is the number of frames, $A$ is the number of antennas, $C$ is the number of chirps per frame, and $N$ is the number of samples per chirp. For each frame $f \in \{1, \ldots, F\}$, antenna $a \in \{1, \ldots, A\}$, and chirp $c \in \{1, \ldots, C\}$, the raw \gls{iq} vector $\mathbf{r}_{f,a,c} \in \mathbb{C}^N$ is first processed as follows:

\subsubsection{\gls{dc} Bias Removal}
The \gls{dc} component of a signal is its average (mean) value over time, corresponding to the zero-frequency part in the frequency domain. In radar signal processing, it often arises from hardware imperfections and contributes no useful information about targets. Removing the \gls{dc} component centers the signal around zero, which improves the accuracy of frequency analysis (e.g., \gls{fft}) and prevents distortion in low-frequency bins. We remove the mean across samples for each chirp to eliminate \gls{dc} offset:

\begin{equation}
    \tilde{\mathbf{r}}_{f,a,c} = \mathbf{r}_{f,a,c} - \frac{1}{N} \sum_{n=1}^{N} \mathbf{r}_{f,a,c}[n].
\end{equation}

\subsubsection{\gls{fft} Computation}
\gls{fft} is used to convert the radar signal from the time domain into the frequency domain. In the time domain, the raw \gls{iq} samples represent signal changes over time within each chirp. However, most useful radar information—such as target distance, motion, and reflection characteristics—is encoded in frequency content. Specifically, in \gls{fmcw} radars, different target ranges correspond to different beat frequencies, so analyzing the frequency spectrum allows us to resolve targets at different distances. The \gls{fft} efficiently decomposes the signal into its constituent frequency components, enabling us to identify and analyze these frequency shifts with high precision. Consequently, an \gls{fft} operation is applied along the sample axis to obtain the frequency-domain representation:

\begin{equation}
    \mathbf{R}_{f,a,c} = \frac{1}{N} \cdot \text{FFT}\left(\tilde{\mathbf{r}}_{f,a,c}\right).
\end{equation}

\subsubsection{Real-valued Frequency Selection}
After computing the \gls{fft}, only the real-valued frequency components are retained because the input radar signals are real-valued (due to the type of the radar which is used in the experiments, i.e. Infineon BGT60TR13C\footnote{\url{https://www.infineon.com/cms/en/product/evaluation-boards/demo-bgt60tr13c}}, and the \gls{fft} of such signals is symmetric about the center.
This means the imaginary frequency component parts contain no additional information—they mirror the real-valued side.
By discarding the redundant half, we reduce computation and focus on the meaningful part of the spectrum. Additionally, energy scaling (typically multiplying by \num{2}) is also applied to preserve the total signal energy when halving the spectrum. This ensures that the amplitude values remain physically meaningful, especially when later used for calibration or signal interpretation:

\begin{equation}
\mathbf{R}^{+}_{f,a,c} = 2 \cdot \mathbf{R}_{f,a,c}[1 : N/2].
\end{equation}

\subsubsection{\gls{ap} Calculation}

Then, the amplitude is calculated by taking the magnitude of the complex spectrum:

\begin{equation}
\mathbf{A}_{f,a,c} = |\mathbf{R}^{+}_{f,a,c}| \in \mathbb{R}^{N/2}.
\end{equation}

\subsubsection{Chirp Averaging}

Finally, an average across all chirps is calculated to obtain the \textbf{\gls{ap}} for each frame and antenna:

\begin{equation}
\bar{\mathbf{A}}_{f,a} = \frac{1}{C} \sum_{c=1}^{C} \mathbf{A}_{f,a,c}
\end{equation}

These \glspl{ap} $\bar{\mathbf{A}}_{f,a} \in \mathbb{R}^{N/2}$ serve as input for the calibration model. The full set of amplitude displacements is stored in a tensor.

\subsection{Temperature-Amplitude Calibration Model}
To compensate for temperature-dependent amplitude variations, a model is developed that predicts and corrects the effect of temperature on the \glspl{ap}. Let $T_f \in \mathbb{R}$ denote the internal radar temperature measured for frame $f$, and $\bar{A}_{f,a,b}$ denote the amplitude value at \gls{fft} bin $b$ for antenna $a$. We assume that the amplitude at a given bin can be approximated as a function of temperature in the training (calibration) phase:

\begin{equation}
\bar{A}_{f,a,b} \approx \mathcal{M}_{a,b}(T_f),
\end{equation}

\noindent where $\mathcal{M}_{a,b}(\cdot)$ denotes a learned temperature-to-amplitude mapping function, trained using amplitude and temperature data from the training phase. This function can be implemented using any regression or \gls{ml} model, such as linear regression, kernel ridge regression, or neural networks.

Assume $T_{\text{REF}}$ is the reference temperature, set as the mean of training temperatures, $\hat{A}^{\text{REF}} = \mathcal{M}_{a,b}(T_{\text{REF}})$ is the predicted amplitude at reference temperature, $\hat{A}^{f} = \mathcal{M}_{a,b}(T_f)$ is the predicted amplitude at frame $f$'s temperature.

A multiplicative correction factor is applied to compensate for the gain change:

\begin{equation}
\tilde{A}_{f,a,b} = \bar{A}_{f,a,b} \cdot \frac{\hat{A}^{\text{REF}}}{\hat{A}^{f}}
\end{equation}

This yields the \textbf{\gls{tcap}}:

\begin{equation}
\tilde{\mathbf{A}}_{f,a} = \bar{\mathbf{A}}_{f,a} \odot \frac{\mathcal{M}_{a} (T_{\text{REF}})}{\mathcal{M}_{a}(T_f)},
\end{equation}

\noindent where $\odot$ denotes element-wise multiplication.

\subsection{Online Training and Inference Phases}

As shown in Fig.\ref{fig:system_model}, during the \textbf{Online Training Phase}, a subset of radar data is used to compute \glspl{ap} and corresponding temperatures. These are used to train the temperature-amplitude model $\mathcal{M}_{a,b}$ for each antenna and \gls{fft} bin of interest. During the \textbf{Inference Phase}, incoming radar data is preprocessed similarly, and temperature measurements are used to compute correction factors using the trained model, yielding \glspl{tcap}.

\subsection{Experimental Setup}
\begin{figure}
\centering
\begin{subfigure}{.49\linewidth}
  \centering
  \includegraphics[width=1\linewidth]{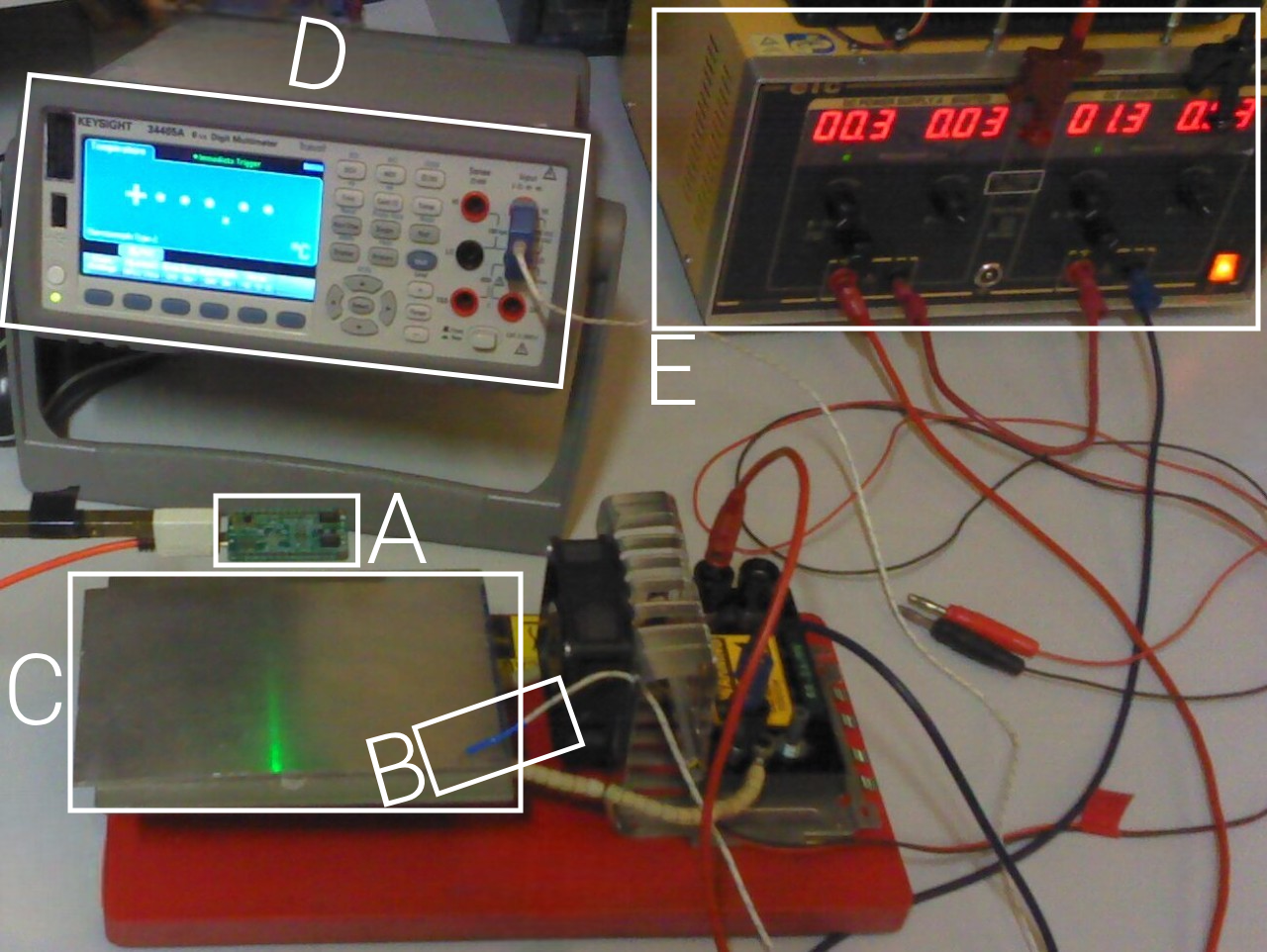}
  \caption{Experiment setup using RGB camera}
  \label{fig:rgb}
\end{subfigure}%
\hfill
\begin{subfigure}{.49\linewidth}
  \centering
  \includegraphics[width=1\linewidth]{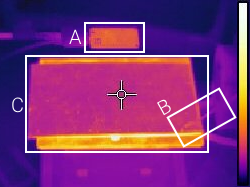}
  \caption{Thermal image of the experiment setup}
  \label{fig:thermal}
\end{subfigure}%
\caption{Experimental setup: A) Infineon BGT60TR13C mmWave radar, B) Thermometer on metallic target, C) Metallic target, D) Keysight temperature reading device, E) Temperature control device.}
\label{fig:experiment_setup}
\end{figure}

The experimental setup, depicted in Fig.\ref{fig:experiment_setup} (with rgb and thermal camera), was designed to evaluate the radar's response to temperature variations in a controlled environment. The setup comprised the following components:
\begin{itemize}
    \item[\textbf{A}:] A mmWave \gls{fmcw} radar, specifically, we utilise the Infineon BGT60TR13C, positioned to transmit and receive signals. The radar has an internal temperature sensor which is used for the callibration purposes.
    \item[\textbf{B}:] A thermometer attached to a metallic target, enabling precise monitoring of the target's temperature during experiments.
    \item[\textbf{C}:] The metallic target, placed within \qty{20}{\centi\meter} of the radar, serving as a potential object of interest for temperature-induced signal analysis. In future work, we plan to explore the possibility of estimating the target's temperature using the calibration mechanism introduced in this paper. To support this, we have already created a setup that allows us to vary the temperature of a potential target and study its effect on the radar signal.
    \item[\textbf{D}:] A Keysight device configured to read temperature measurements from the thermometer, providing accurate data on the target's temperature.
    \item[\textbf{E}:] A temperature control device utilizing electric current to adjust the temperature of the metallic target, facilitating dynamic thermal variations.
\end{itemize}

The experiment was conducted across two distinct settings to isolate the effects of temperature on radar performance:

\begin{enumerate}
    \item \textbf{Radar facing Direction 1}: The radar was oriented in a direction that did not face the metallic surface. The target's temperature was maintained at room temperature, with no moving targets present. Internal radar temperature variations were induced by both internal changes and external influences, such as a fan altering ambient conditions.
    \item \textbf{Radar facing Direction 2}: The radar was oriented in a different direction that also did not face the metallic surface. Similar to the first setting, the target's temperature remained at room temperature, and no moving targets were present. Internal radar temperature fluctuations were again influenced by internal dynamics and external fan-induced changes.
\end{enumerate}

In all settings, the experimental environment was controlled to exclude moving targets or additional variables, ensuring that temperature effects on the radar signal could be isolated and analyzed effectively.

In this study, a linear regression model is utilized to predict amplitude based on internal radar temperature, offering a straightforward yet effective approach to modeling their linear relationship. The dataset is divided with \qty{70}{\percent} of the timesteps allocated for training to capture the temperature-amplitude dynamics, and the remaining \qty{30}{\percent} used for testing to validate the model’s performance. The radar operates within a frequency range from a start frequency of \qty{58}{\giga\hertz} to an end frequency of \qty{63.5}{\giga\hertz}, equipped with three antennas, and processes data across two chirps per frame with \num{32} samples per chirp. The first setting comprises \qty{34.7}{\kilo{}} frames, while the second setting includes \qty{27.5}{\kilo{}} frames, providing a robust dataset for comparative analysis across the experimental conditions.

\section{Results and Discussion}
\begin{figure*}
    \centering
    \begin{subfigure}{.333\textwidth}
      \centering
      \includegraphics[width=1\linewidth]{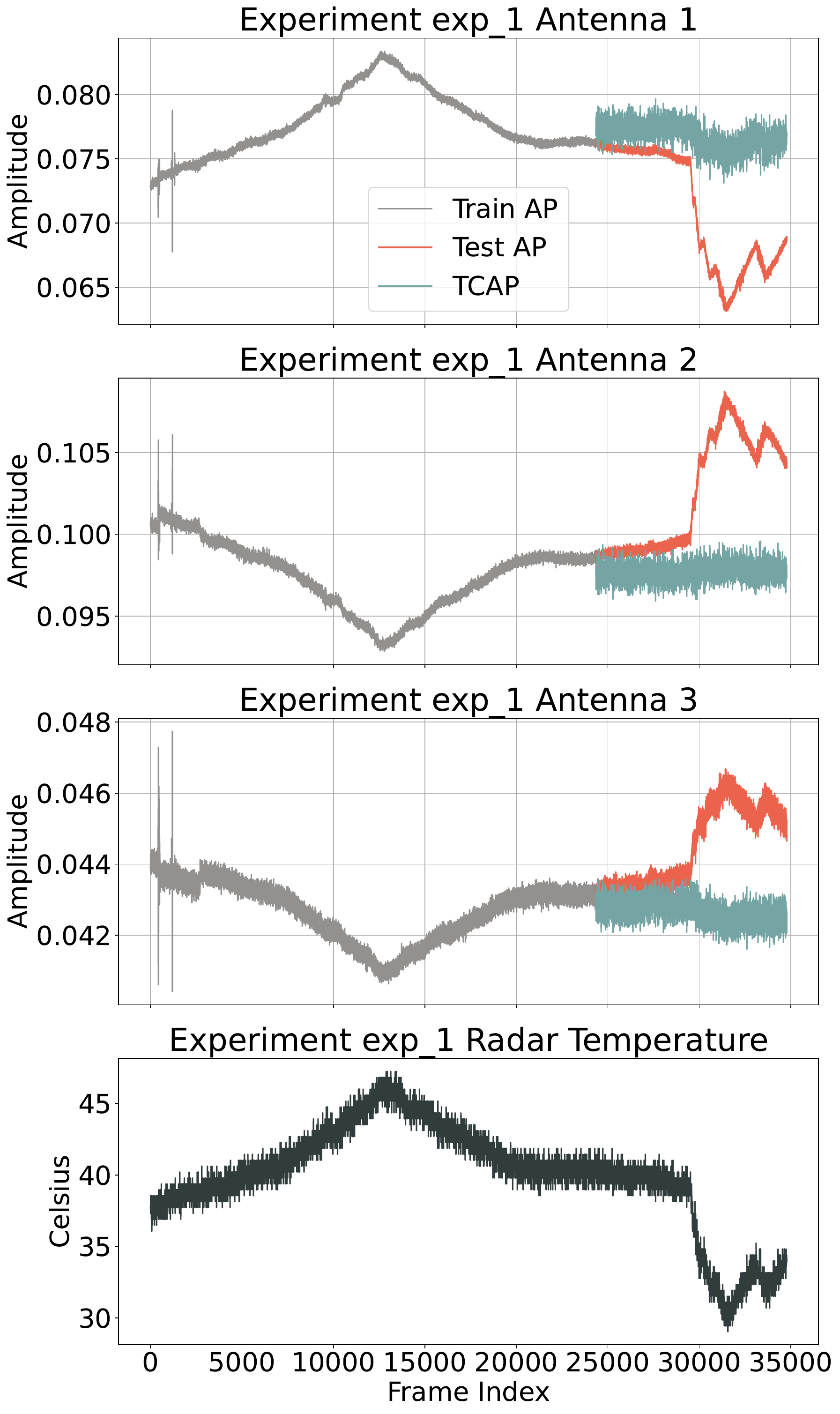}
      \caption{First experiment setting}
      \label{fig:first_setting_results}
    \end{subfigure}%
    \hfill
    \begin{subfigure}{.333\textwidth}
      \centering
      \includegraphics[width=1\linewidth]{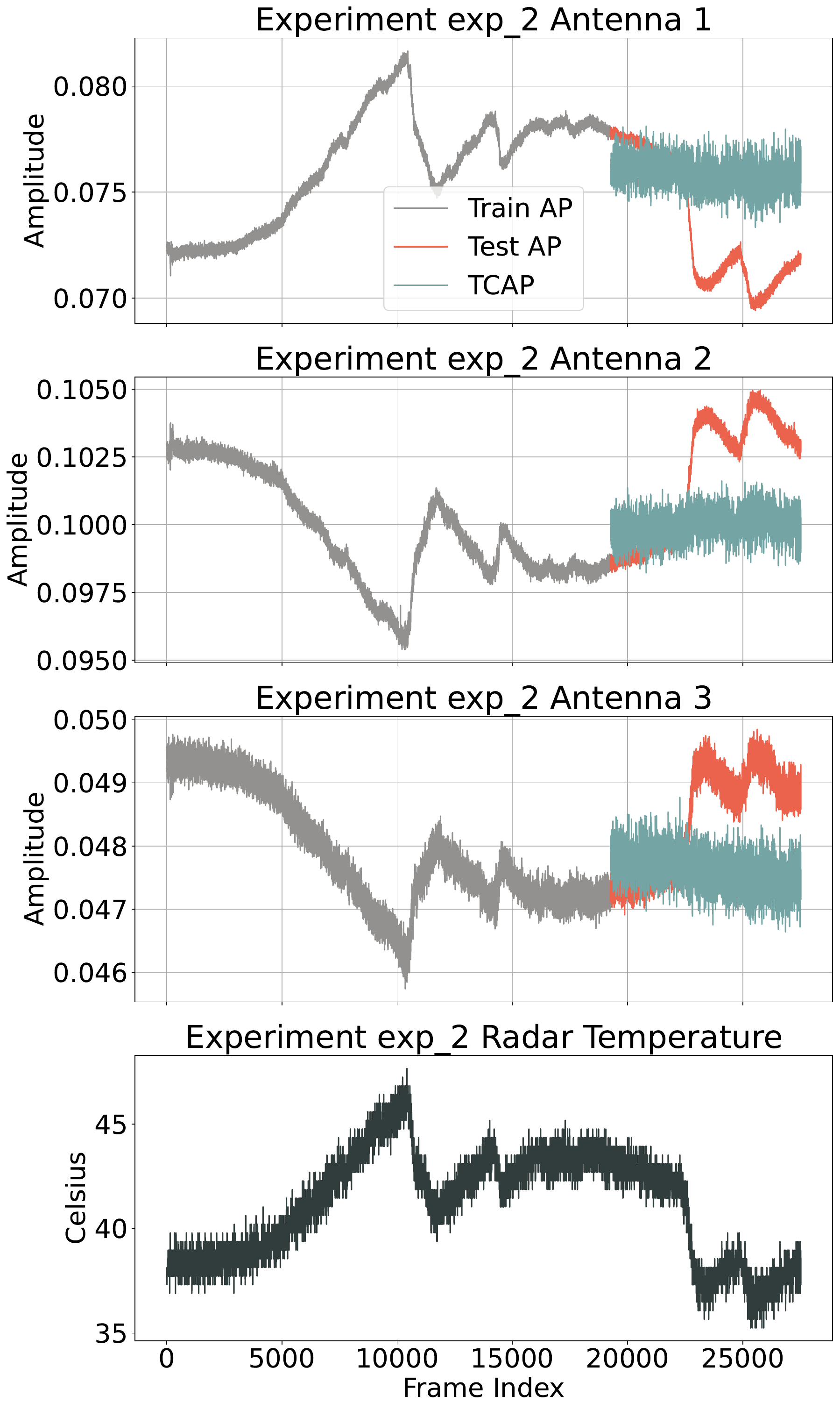}
      \caption{Second experiment setting}
      \label{fig:second_setting_results}
    \end{subfigure}%
    \hfill
    \begin{subfigure}{.333\textwidth}
      \centering
      \includegraphics[width=1\linewidth]{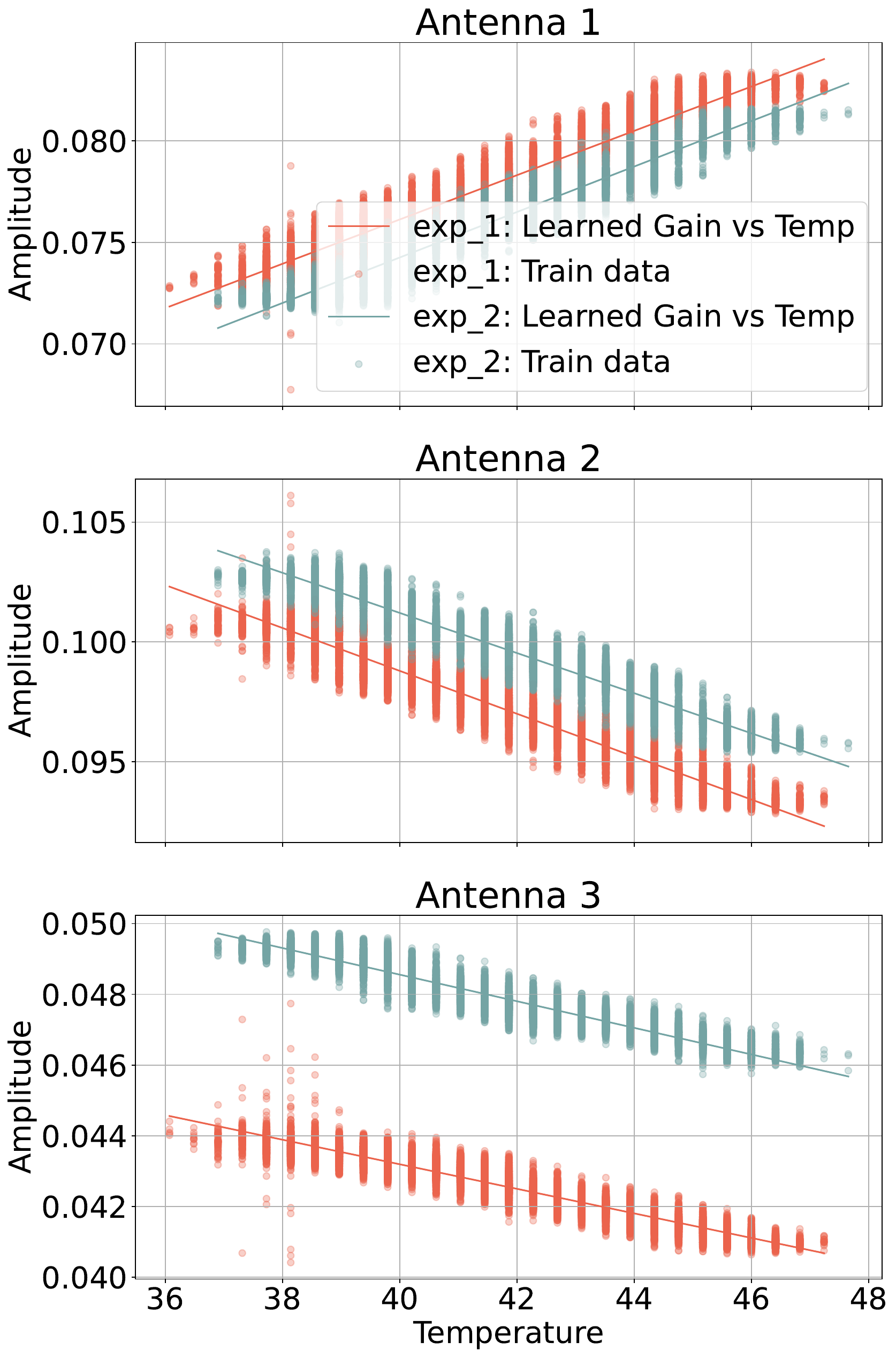}
      \caption{Learned model for both settings}
      \label{fig:linear_modeling}
    \end{subfigure}%
    \caption{Comparison of experimental results across two settings. (a) \glspl{ap} for training/testing phases and \glspl{tcap} for varying receive antennas, and internal radar temperature in the first setting. (b) \glspl{ap} for training/testing phases and \glspl{tcap} for varying receive antennas, and internal radar temperature in the second setting. (c) Linear regression models for predicting amplitude based on internal radar temperature across different antennas in both settings, distinguished by color.}
    \label{fig:results_figure}
\end{figure*}

The experimental data reveals insights into temperature-dependent amplitude behavior across two settings for three antennas. The radar temperature fluctuates between \qtyrange{30}{45}{\degreeCelsius}, strongly correlating with the amplitude shifts in both settings as shown in Fig.\ref{fig:results_figure}. In the first setting shown in Fig.\ref{fig:first_setting_results}, Antenna 1 exhibits an unstable amplitude around \num{0.075}, with Train \gls{ap} and Test \gls{ap} fluctuating aligned with the internal temperature changes while \gls{tcap} showing a stable amplitude, suggesting effective temperature compensation. A similar trend is observed for both Antenna 2 and 3 with a stable \gls{tcap} highlighting the effectiveness of the proposed scheme.

In the second setting shown in Fig.\ref{fig:second_setting_results}, Antenna 1’s amplitude rises to \num{0.080}, with Test \gls{ap} showing greater variability, and \gls{tcap} smoothing this trend. Antenna 2’s amplitude peaks at \num{0.1025}, with a jagged Test \gls{ap} suggesting measurement instability due to the internal temperature change, while \gls{tcap} reduces this noise. Antenna 3 maintains a low amplitude (around \num{0.049}), with Test \gls{ap} exhibiting a sharp decline, and \gls{tcap} offering modest correction.

The learned models indicate a consistent negative slope for Antennas 2 and 3 across settings, with Antenna 1 showing a positive slope, reflecting divergent thermal responses. The Pearson correlations shown in Table~\ref{tab:pearson_correlation_results} underscore these trends, with Antenna 1’s strong positive \gls{ap} correlation contrasting with the strong negative correlations for Antennas 2 and 3, and \gls{tcap} correlations highlighting residual temperature effects.

The results indicate a clear dependency of \glspl{ap} on internal radar temperature, with distinct behaviors across antennas and settings. The strong correlations between temperature and \gls{ap} suggest that temperature variations significantly influence raw amplitude measurements, necessitating compensation strategies. The moderate correlations with \gls{tcap} reflect the effectiveness of temperature compensation, though residual effects are evident, particularly for Antennas 1 and 3 in the first setting.

The learned linear regression models show similar slopes across both settings, suggesting a consistent underlying physical relationship between temperature and amplitude driven by the radar's thermal response and antenna characteristics, despite differing intercepts. These intercepts likely reflect setting-specific baseline amplitudes due to variations in initial calibration, environmental factors, or hardware configurations. The negative correlations for Antennas 2 and 3 indicate an inverse temperature-amplitude relationship, possibly from thermal expansion or material properties affecting signal attenuation differently than Antenna 1’s positive correlation, with slope consistency reinforcing the model's robustness, though intercept differences necessitate setting-specific adjustments, warranting future controlled experiments to isolate these variations.

Although the internal calibration framework for mmWave FMCW radars is setting-dependent, many practical applications, such as structural health monitoring and arterial pulse detection, benefit from relatively stable environmental and operational conditions. This stability allows for an effective initial calibration phase at deployment, as the setting does not change rapidly. By performing calibration at the outset, the framework can accurately compensate for temperature-induced drifts, ensuring consistent and reliable radar performance over extended periods without the need for frequent recalibration.

\begin{table}
\centering
\caption{Pearson correlation between the internal temperature of the radar with \gls{ap} and \gls{tcap}.}
\label{tab:pearson_correlation_results}
\begin{tabular}{@{}cccc@{}}
\toprule
Setting & Antenna & $\operatorname{PR}(T, AP)$ & $\operatorname{PR}(T, TCAP)$ \\ \midrule
\rowcolor{lightgray!60} First & 1 & 0.99 & 0.64 \\
 \cellcolor{lightgray!60} & 2 & -0.99 & -0.16 \\
 \rowcolor{lightgray!60} & 3 & -0.98 & 0.64 \\
Second & 1 & 0.98 & 0.19 \\
 \rowcolor{lightgray!60} \cellcolor{white!60}& 2 & -0.98 & -0.17 \\
 & 3 & -0.96 & 0.58 \\\bottomrule
\vspace{0.1em}
\end{tabular}%
\\
\raggedright
{\footnotesize $\operatorname{PR}(.)$ is the pearson correlation operator, $T$ is the internal temperature of the radar, \gls{ap} is the amplitude profile before compensation, and \gls{tcap} is the temperature compensated amplitude profile.}
\end{table}

Moreover, the temperature of the BGT60TR16C radar chip was estimated using the formula $\text{Temp} = \frac{\text{Tsense} - a}{b}$, where $\text{Tsense}$ is the temperature sensor readout in volts, and $a$ and $b$ are the temperature sensor offset and slope, respectively, as specified in the device datasheet~\footnote{https://www.infineon.com/assets/row/public/documents/24/49/infineon-bgt60tr13c-datasheet-en.pdf}. This calibration approach allows conversion of the voltage readout into an accurate temperature value expressed in Celsius. To improve long-term stability and measurement accuracy, the \gls{tsr} provided by the radar chip can be used in a differential measurement scheme. These enhancements offer a valuable opportunity for improving calibration stability and reducing sensor drift in future work. Importantly, they build upon the core calibration concept established in this study.

\section{Conclusion}
This study presents a novel internal calibration framework for mmWave \gls{fmcw} radars, effectively mitigating temperature-induced amplitude drifts to enhance measurement reliability in dense wireless networks. Our experimental results demonstrate that the proposed temperature compensation model significantly reduces the Pearson correlation between internal radar temperature and \glspl{ap} across multiple antennas and settings, showcasing robust performance under varying thermal conditions. This framework ensures scalability and minimal computational overhead, making it suitable for resource-constrained environments.

The contributions of this research advance the field of radar signal processing by introducing a privacy-scalable, ethical calibration approach that relies solely on internal sensor data, eliminating the need for external references that could compromise user privacy. This is particularly impactful for applications like non-invasive glucose monitoring, where precise and reliable measurements are critical, and privacy concerns are paramount. The framework's ability to maintain accuracy across a temperature range of \qtyrange{30}{45}{\degreeCelsius} underscores its practical utility in real-world scenarios, such as biomedical sensing, structural health monitoring, and autonomous vehicle systems.

Potential applications of this method extend to any mmWave \gls{fmcw} radar-based system requiring robust performance in dynamic thermal environments. In healthcare, it supports non-invasive monitoring of physiological parameters, such as glucose levels or arterial pulse detection. In infrastructure, it enhances the reliability of structural health assessments, while in automotive systems, it improves object detection accuracy under varying operational conditions.

Future research directions include exploring advanced machine learning models, such as neural networks, to capture non-linear temperature-amplitude relationships, potentially improving compensation accuracy. In addition to compensation, we also aim to leverage the proposed calibration mechanism to estimate the temperature of the target itself. To this end, we have developed an experimental setup that allows controlled variation of a target’s temperature to study its impact on the radar signal. Extending the framework to account for external environmental factors, such as humidity and pressure, could further enhance robustness. Moreover, investigating the framework’s performance in multi-radar dense network scenarios and optimizing its computational efficiency for deployment on low-power devices remain promising avenues for further development.

\balance

\section*{Acknowledgment}
We acknowledge funding by the European Union through the Horizon Europe EIC Pathfinder projects SUSTAIN (Grant Agreement No. 101071179) and HOLDEN (Grant Agreement No. 101099491). Views and opinions expressed are those of the authors only and do not necessarily reflect those of the European Union or the European Innovation Council and SMEs Executive Agency (EISMEA). Neither the European Union nor the granting authority can be held responsible for them.

\bibliographystyle{ACM-Reference-Format}
\bibliography{references.bib}

%%% -*-BibTeX-*-
%%% Do NOT edit. File created by BibTeX with style
%%% ACM-Reference-Format-Journals [18-Jan-2012].

\begin{thebibliography}{14}

%%% ====================================================================
%%% NOTE TO THE USER: you can override these defaults by providing
%%% customized versions of any of these macros before the \bibliography
%%% command.  Each of them MUST provide its own final punctuation,
%%% except for \shownote{} and \showURL{}.  The latter two
%%% do not use final punctuation, in order to avoid confusing it with
%%% the Web address.
%%%
%%% To suppress output of a particular field, define its macro to expand
%%% to an empty string, or better, \unskip, like this:
%%%
%%% \newcommand{\showURL}[1]{\unskip}   % LaTeX syntax
%%%
%%% \def \showURL #1{\unskip}           % plain TeX syntax
%%%
%%% ====================================================================

\ifx \showCODEN    \undefined \def \showCODEN     #1{\unskip}     \fi
\ifx \showISBNx    \undefined \def \showISBNx     #1{\unskip}     \fi
\ifx \showISBNxiii \undefined \def \showISBNxiii  #1{\unskip}     \fi
\ifx \showISSN     \undefined \def \showISSN      #1{\unskip}     \fi
\ifx \showLCCN     \undefined \def \showLCCN      #1{\unskip}     \fi
\ifx \shownote     \undefined \def \shownote      #1{#1}          \fi
\ifx \showarticletitle \undefined \def \showarticletitle #1{#1}   \fi
\ifx \showURL      \undefined \def \showURL       {\relax}        \fi
% The following commands are used for tagged output and should be
% invisible to TeX
\providecommand\bibfield[2]{#2}
\providecommand\bibinfo[2]{#2}
\providecommand\natexlab[1]{#1}
\providecommand\showeprint[2][]{arXiv:#2}

\bibitem[Baer et~al\mbox{.}(2014)]%
        {baer2014contactless}
\bibfield{author}{\bibinfo{person}{Christoph Baer}, \bibinfo{person}{Timo Jaeschke}, \bibinfo{person}{Nils Pohl}, {and} \bibinfo{person}{Thomas Musch}.} \bibinfo{year}{2014}\natexlab{}.
\newblock \showarticletitle{Contactless detection of state parameter fluctuations of gaseous media based on an mm-wave FMCW radar}.
\newblock \bibinfo{journal}{\emph{IEEE Transactions on Instrumentation and Measurement}} \bibinfo{volume}{64}, \bibinfo{number}{4} (\bibinfo{year}{2014}), \bibinfo{pages}{865--872}.
\newblock


\bibitem[Bahmani et~al\mbox{.}(2025)]%
        {bahmani2025non}
\bibfield{author}{\bibinfo{person}{Nima Bahmani}, \bibinfo{person}{Dariush Salami}, \bibinfo{person}{H{\"u}seyin Yi{\u{g}}itler}, \bibinfo{person}{Juhapekka Hietala}, \bibinfo{person}{Tuukka Panula}, {and} \bibinfo{person}{Stephan Sigg}.} \bibinfo{year}{2025}\natexlab{}.
\newblock \showarticletitle{Non-Invasive Arterial Pulse Detection with Millimeter-wave Radar and Comparison With Photoplethysmography}. In \bibinfo{booktitle}{\emph{Proceedings of the 2025 ACM International Symposium on Wearable Computers}}. ACM.
\newblock


\bibitem[Machado and Mancheno(2018)]%
        {machado2018automotive}
\bibfield{author}{\bibinfo{person}{Sanoal Machado} {and} \bibinfo{person}{Santiago Mancheno}.} \bibinfo{year}{2018}\natexlab{}.
\newblock \showarticletitle{Automotive FMCW radar development and verification methods}.
\newblock  (\bibinfo{year}{2018}).
\newblock


\bibitem[Nikandish et~al\mbox{.}(2024)]%
        {nikandish2024contactless}
\bibfield{author}{\bibinfo{person}{Reza Nikandish}, \bibinfo{person}{Caroline Sheedy}, \bibinfo{person}{Jiayu He}, \bibinfo{person}{Ruth Crowe}, {and} \bibinfo{person}{Deeksha Rao}.} \bibinfo{year}{2024}\natexlab{}.
\newblock \showarticletitle{Contactless Glucose Sensing Using Miniature mm-Wave Radar and Tiny Machine Learning}.
\newblock \bibinfo{journal}{\emph{IEEE Journal of Microwaves}} (\bibinfo{year}{2024}).
\newblock


\bibitem[Omer et~al\mbox{.}(2020)]%
        {omer2020blood}
\bibfield{author}{\bibinfo{person}{Ala~Eldin Omer}, \bibinfo{person}{Safieddin Safavi-Naeini}, \bibinfo{person}{Richard Hughson}, {and} \bibinfo{person}{George Shaker}.} \bibinfo{year}{2020}\natexlab{}.
\newblock \showarticletitle{Blood glucose level monitoring using an FMCW millimeter-wave radar sensor}.
\newblock \bibinfo{journal}{\emph{Remote Sensing}} \bibinfo{volume}{12}, \bibinfo{number}{3} (\bibinfo{year}{2020}), \bibinfo{pages}{385}.
\newblock


\bibitem[Palipana et~al\mbox{.}(2021)]%
        {palipana2021pantomime}
\bibfield{author}{\bibinfo{person}{Sameera Palipana}, \bibinfo{person}{Dariush Salami}, \bibinfo{person}{Luis~A Leiva}, {and} \bibinfo{person}{Stephan Sigg}.} \bibinfo{year}{2021}\natexlab{}.
\newblock \showarticletitle{Pantomime: Mid-air gesture recognition with sparse millimeter-wave radar point clouds}.
\newblock \bibinfo{journal}{\emph{Proceedings of the ACM on interactive, mobile, wearable and ubiquitous technologies}} \bibinfo{volume}{5}, \bibinfo{number}{1} (\bibinfo{year}{2021}), \bibinfo{pages}{1--27}.
\newblock


\bibitem[Salami(2024)]%
        {salami2024spectrum}
\bibfield{author}{\bibinfo{person}{Dariush Salami}.} \bibinfo{year}{2024}\natexlab{}.
\newblock \showarticletitle{Spectrum-aware Human-Centric Sensing (HCS) using mmWave radars}.
\newblock  (\bibinfo{year}{2024}).
\newblock


\bibitem[Salami et~al\mbox{.}(2022)]%
        {salami2022tesla}
\bibfield{author}{\bibinfo{person}{Dariush Salami}, \bibinfo{person}{Ramin Hasibi}, \bibinfo{person}{Sameera Palipana}, \bibinfo{person}{Petar Popovski}, \bibinfo{person}{Tom Michoel}, {and} \bibinfo{person}{Stephan Sigg}.} \bibinfo{year}{2022}\natexlab{}.
\newblock \showarticletitle{Tesla-rapture: A lightweight gesture recognition system from mmwave radar sparse point clouds}.
\newblock \bibinfo{journal}{\emph{IEEE Transactions on Mobile Computing}} \bibinfo{volume}{22}, \bibinfo{number}{8} (\bibinfo{year}{2022}), \bibinfo{pages}{4946--4960}.
\newblock


\bibitem[Salami et~al\mbox{.}(2024)]%
        {salami2024angle}
\bibfield{author}{\bibinfo{person}{Dariush Salami}, \bibinfo{person}{Ramin Hasibi}, \bibinfo{person}{Stefano Savazzi}, \bibinfo{person}{Tom Michoel}, {and} \bibinfo{person}{Stephan Sigg}.} \bibinfo{year}{2024}\natexlab{}.
\newblock \showarticletitle{Angle-Agnostic Radio Frequency Sensing Integrated into 5G-NR}.
\newblock \bibinfo{journal}{\emph{IEEE Sensors Journal}} (\bibinfo{year}{2024}).
\newblock


\bibitem[Salami et~al\mbox{.}(2023)]%
        {salami2023joint}
\bibfield{author}{\bibinfo{person}{Dariush Salami}, \bibinfo{person}{Wanru Ning}, \bibinfo{person}{Kalle Ruttik}, \bibinfo{person}{Riku J{\"a}ntti}, {and} \bibinfo{person}{Stephan Sigg}.} \bibinfo{year}{2023}\natexlab{}.
\newblock \showarticletitle{A joint radar and communication approach for 5G NR using reinforcement learning}.
\newblock \bibinfo{journal}{\emph{IEEE Communications Magazine}} \bibinfo{volume}{61}, \bibinfo{number}{5} (\bibinfo{year}{2023}), \bibinfo{pages}{106--112}.
\newblock


\bibitem[Schuster et~al\mbox{.}(2006)]%
        {schuster2006performance}
\bibfield{author}{\bibinfo{person}{Stefan Schuster}, \bibinfo{person}{Stefan Scheiblhofer}, \bibinfo{person}{Leonhard Reindl}, {and} \bibinfo{person}{Andreas Stelzer}.} \bibinfo{year}{2006}\natexlab{}.
\newblock \showarticletitle{Performance evaluation of algorithms for SAW-based temperature measurement}.
\newblock \bibinfo{journal}{\emph{IEEE transactions on ultrasonics, ferroelectrics, and frequency control}} \bibinfo{volume}{53}, \bibinfo{number}{6} (\bibinfo{year}{2006}), \bibinfo{pages}{1177--1185}.
\newblock


\bibitem[Simon et~al\mbox{.}(2021)]%
        {simon2021experimental}
\bibfield{author}{\bibinfo{person}{Jonas Simon}, \bibinfo{person}{Thomas Maetz}, \bibinfo{person}{Jochen Moll}, \bibinfo{person}{Viktor Krozer}, {and} \bibinfo{person}{Stefan Krause}.} \bibinfo{year}{2021}\natexlab{}.
\newblock \showarticletitle{Experimental results on the influence of temperature and humidity on FMCW radar signals at 60 GHz}. In \bibinfo{booktitle}{\emph{2021 15th European Conference on Antennas and Propagation (EuCAP)}}. IEEE, \bibinfo{pages}{1--4}.
\newblock


\bibitem[Toledo et~al\mbox{.}(2020)]%
        {toledo2020absolute}
\bibfield{author}{\bibinfo{person}{Felipe Toledo}, \bibinfo{person}{Julien Delano{\"e}}, \bibinfo{person}{Martial Haeffelin}, \bibinfo{person}{Jean-Charles Dupont}, \bibinfo{person}{Susana Jorquera}, {and} \bibinfo{person}{Christophe Le~Gac}.} \bibinfo{year}{2020}\natexlab{}.
\newblock \showarticletitle{Absolute calibration method for frequency-modulated continuous wave (FMCW) cloud radars based on corner reflectors}.
\newblock \bibinfo{journal}{\emph{Atmospheric Measurement Techniques}} \bibinfo{volume}{13}, \bibinfo{number}{12} (\bibinfo{year}{2020}), \bibinfo{pages}{6853--6875}.
\newblock


\bibitem[Wang et~al\mbox{.}(2024)]%
        {wang2024influences}
\bibfield{author}{\bibinfo{person}{Qiupin Wang}, \bibinfo{person}{Guangqiong Xia}, \bibinfo{person}{Yingke Xie}, \bibinfo{person}{Pu Ou}, \bibinfo{person}{Chaotao He}, \bibinfo{person}{Shan Hu}, \bibinfo{person}{Fengling Zhang}, \bibinfo{person}{Maorong Zhao}, {and} \bibinfo{person}{Zhengmao Wu}.} \bibinfo{year}{2024}\natexlab{}.
\newblock \showarticletitle{Influences of Thermal Effect on the Performance of FMCW Signal Generated by Current-Modulated DFB-LDs}.
\newblock \bibinfo{journal}{\emph{IEEE Journal of Quantum Electronics}} (\bibinfo{year}{2024}).
\newblock


\end{thebibliography}
\end{document}